\documentclass[%
 aip,
apl,
 sd,%
 amsmath,amssymb,
 reprint,
]{revtex4-1}

\usepackage{graphicx}% Include figure files
\usepackage{dcolumn}% Align table columns on decimal point
\usepackage{bm}% bold math
\usepackage{hyperref}
\hypersetup{
    colorlinks=true,
    citecolor = blue,
    linkcolor=blue,
    filecolor=blue,      
    urlcolor=blue,
}
\begin{document}

\preprint{AIP/123-QED}

\title[Printing wet-on-wet: attraction and repulsion of drops on a viscous film]{Printing wet-on-wet: attraction and repulsion of drops on a viscous film}

\author{M. A. Hack}
\email{m.a.hack@utwente.nl.}
\affiliation{Physics of Fluids Group, Faculty of Science and Technology, University of Twente, P.O. Box 217, 7500 AE Enschede, The Netherlands}
\author{M. Costalonga}
\affiliation{Physics of Fluids Group, Faculty of Science and Technology, University of Twente, P.O. Box 217, 7500 AE Enschede, The Netherlands}
\author{T. Segers}
\affiliation{Physics of Fluids Group, Faculty of Science and Technology, University of Twente, P.O. Box 217, 7500 AE Enschede, The Netherlands}
\author{S. Karpitschka}
\affiliation{Max Planck Institute for Dynamics and Self-Organization, Am Fa{\ss}berg 17, 37077 G\"ottingen, Germany}
\author{H. Wijshoff}
\affiliation{Department of Mechanical Engineering, Eindhoven University of Technology, P.O. Box 513, 5600 MB Eindhoven, The Netherlands}
\affiliation{Oc\'e Technologies B.V., P.O. Box 101, 5900 MA Venlo, The Netherlands}
\author{J. H. Snoeijer}
\affiliation{Physics of Fluids Group, Faculty of Science and Technology, University of Twente, P.O. Box 217, 7500 AE Enschede, The Netherlands}

\date{\today}

\begin{abstract}
Wet-on-wet printing is frequently used in inkjet printing for graphical and industrial applications, where substrates can be coated with a thin liquid film prior to ink drop deposition. Two drops placed close together are expected to interact via deformations of the thin viscous film, but the nature of these capillary interactions is unknown. Here we show that the interaction can be attractive or repulsive depending on the distance separating the two drops. The distance at which the interaction changes from attraction to repulsion is found to depend on the thickness of the film, and increases over time. We reveal the origin of the non-monotonic interactions, which lies in the appearance of a visco-capillary wave on the thin film induced by the drops. Using the thin-film equation we identify the scaling law for the spreading of the waves, and demonstrate that this governs the range over which interaction is observed.
\end{abstract}


\maketitle

Solid particles at a liquid-gas interface have a tendency to form clusters due to capillarity-driven interactions. This phenomenon is known as the ``Cheerios effect", named after the floating cereals that form clusters at the milk-air interface.\cite{Vella_AJP_2005} Manifestations of the Cheerios effect are also found in biology. For instance, mosquito eggs aggregate on the surface of a pond to form rafts.\cite{Loudet_EPJ_2011} Capillary interactions have also been observed between liquid drops\cite{Karpitschka_PNAS_2016, Pandey_SM_2017} and between solid particles\cite{Chakrabarti_EPL_2015} on soft gels. Soft gels are solids but share many properties, such as having a surface tension, with highly viscous liquids.\cite{Snoeijer_PRF_2016} Capillary interactions are also relevant to technological applications, which range from self-assembly\cite{Bowden_S_1997, Bowden_JACS_1999, Bowden_JPC_2000, Wolfe_L_2003} to drop condensation.\cite{Sokuler_L_2010}

In an industrial setting, capillary driven drop interactions play an important role in inkjet printing. Substrates are frequently covered by a first layer of ink before a second ink layer is applied, or can be coated with a thin liquid primer layer prior to ink deposition.\cite{Wijshoff_COCIS_2018} Such primer layers contain salts that destabilize the colloidal pigment particles and thereby increase their sedimentation rate, which enhances print quality.\cite{Oko_CSPEA_2014} Typically, the primer layer thickness is similar to the size of the ink drops, since both are deposited using a similar printhead.\cite{Varela_CEP_2011} We have observed interaction between ink drops deposited on such a primer layer. However, the nature of the capillary interactions between drops deposited on a thin liquid film is still poorly understood. 

%%%%% FIG 1
\begin{figure}
\includegraphics{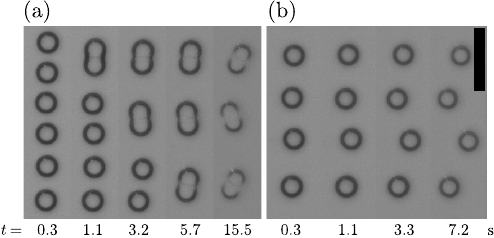}
\caption{\label{fig:fig1}Interaction of water drops (radius $R = 45$ $\mu$m) printed on a thin silicone oil film (thickness $h_0 = 5.7$ $\mu$m). Two types of interactions are observed: (a) Attractive interaction causes the drops to form drop pairs. See also Movie S1. (b) Repulsive interaction results in a zigzag-like pattern of drops. See also Movie S2. The difference between the two experiments is the initial distance between the drops in the printed line. The scale bar represents 400 $\mu$m.}
\end{figure}
%%%%%

%%%%% FIG 2
\begin{figure*}
\includegraphics{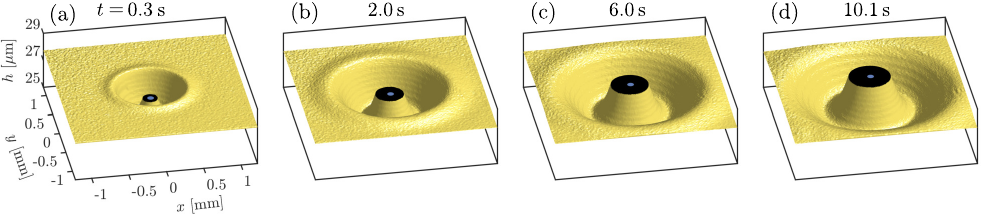}
\caption{\label{fig:fig2}(Color online) Evolution of the profile of the viscous film ($h_0=28$ $\mu$m) around a drop ($R = 45$ $\mu$m) located at the origin. The film exhibits a wavelike deformation that broadens over time: (a) $t=0.3$ s, (b) $t=2$ s, (c) $t=6$ s, (d) $t=10$ s. The axes in (a) also apply to (b)--(d). The black circle in the center corresponds to a region where digital holographic microscopy cannot properly resolve the film's surface, the blue circle denotes the drop's diameter and position in the $xy$-plane.}
\end{figure*}
%%%%%

In this Letter we experimentally study capillary interactions between drops on thin liquid films. We focus on the case where the drops and films are immiscible, which eliminates mixing and Marangoni effects and isolates the Cheerios-like interactions. The essence of our experiment is shown in Fig.~\ref{fig:fig1}: a row of water drops (MilliQ, Millipore Corporation) with a radius of $R$~=~45~$\mu$m (which is the same in all experiments) is jetted onto a thin silicone oil film (Basildon Chemical Co. Ltd.) with a thickness $h_0 =$ 5.7~$\mu$m and viscosity $\eta_\textrm{o}$~=~1~Pa$\cdot$s using a piezo-driven pipette (AD-K-501, Microdrop Technologies). We observe both attractive (Fig.~\ref{fig:fig1}a) and repulsive (Fig.~\ref{fig:fig1}b) drop interactions, where the only difference between these experiments is the distance between the jetted drops. Attraction, as shown in Fig.~\ref{fig:fig1}a, results in drop pairs. The entrainment of a thin oil film between the drops delays their coalescence.\cite{Boreyko_PNAS_2014} In Fig.~\ref{fig:fig1}b, by contrast, the drops are pushed out of the initially straight line, resulting in a zigzag-like configuration. The increase in drop distance clearly points to a repulsive interaction. The semi-coalesced drop pairs in Fig.~\ref{fig:fig1}a also show a zigzag structure at $t~=~15.5$~s, which indicates a possible repulsive interaction for $t~>~3.2$ s. Hence, we find that drops on thin viscous films exhibit intricate non-monotonic interactions.  

The non-monotonic nature of the interactions can be traced back to the surface deformation induced by a single drop. In Fig.~\ref{fig:fig2} we show the profile of the viscous film ($h=28$~$\mu$m) at various times after drop deposition, measured using digital holographic microscopy\cite{Kim_SR_2010} (abbreviated DHM, R-1000, Lync\'ee Tec). The measured surface deformations are non-monotonic and extend over a distance of approximately 1~mm, almost two orders of magnitude larger than the size of the drop. The wave-like profile results from volume conversation: liquid is pulled up to create a meniscus close to the drop, and a capillary wave connects the meniscus to the flat film far away from the drop. The perturbed profile of the liquid film continues to broaden over time as is shown in Fig.~\ref{fig:fig2}a--d. 

Since the interactions between drops are induced by perturbations of the viscous film, we expect the range of the interaction to increase over time. Such a time-dependence in the interaction law is fundamentally different from the usual Cheerios effect (particles at an interface of a deep pool \cite{Vella_AJP_2005}) or the ``inverted Cheerios effect" (drops\cite{Karpitschka_PNAS_2016, Pandey_SM_2017} or particles \cite{Chakrabarti_L_2014, Chakrabarti_EPL_2015, Pandey_EL_2018} on elastic layers). In those cases, the  deformation by a single particle reaches a steady state, so the interaction law is constant over time. In the present case, by contrast, the time-scale of the change in the deformation of the film is similar to the time-scale of the induced drop motion. This makes it very challenging to quantify the detailed interaction law. For this reason we focus on finding the (time-dependent) range of attractive interaction, and correlate this to the evolution of the profile of the viscous film.

\begin{figure}
\includegraphics{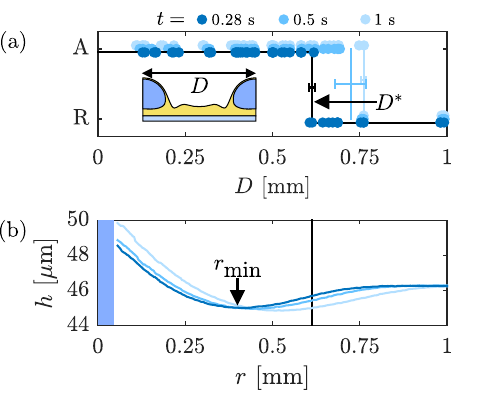}
\caption{\label{fig:fig3}(Color online) (a) Interaction type of two drops separated by a distance $D$ (A = attraction, R = repulsion), at $t = 0.28$,  0.5, 1~s (film thickness $h_0 = 46$ $\mu$m). The distance $D$ is measured center-to-center (inset). The arrow indicates $D^\textrm{*}$ for $t = 0.28$~s. The horizontal bars indicate the error in $D^\textrm{*}$. (b) The surface profile induced by a single drop (same conditions as panel (a)). The black line indicates the position of $D^\textrm{*}$ at $t = 0.28$ s, while the arrow indicates $r_\textrm{min}$. The blue region close to $r = 0$ indicates the radius of the drop.}
\end{figure}

To reveal how the interaction range depends on time and film thickness, we focus on the case of two drops, as sketched in the inset of Fig.~\ref{fig:fig3}a. Oil films of initially uniform thickness $h_0$ were spin-coated on a hydrophobic glass microscope slide (Menzel-Gl\"aser), $h_0$ was varied by changing the rotational speed and spinning time of the spin-coater. The hydrophobization, performed by vapor deposition of trichloro(octadecyl)silane (Sigma-Aldrich), resulted in a contact angle of water on glass sufficiently large to prevent rupture of the silicone oil film underneath the drops through `rewetting'.\cite{Carlson_EPL_2013} The thickness of the silicone oil films was measured using reflectometry (HR2000+ spectrometer with HL-2000-FHSA halogen light source, Ocean Optics).\cite{Reizman_JAP_1965} The interfacial tension between the water drops, with surface tension $\gamma_\text{w}$~=~72~mN~m$^{-1}$, and the silicone oil film, with surface tension $\gamma_\text{o}$~=~21.2~mN~m$^{-1}$, was $\gamma_\text{wo} \approx 20$~mN m$^{-1}$. Consequently, since $\gamma_\text{wo}~+~\gamma_\text{o}~<~\gamma_\text{w}$, a thin silicone oil film engulfed the water drops.\cite{Carlson_EPL_2013, Boreyko_PNAS_2014, Smith_SM_2013, Schellenberger_SM_2015} The oil-coated glass substrate was mounted on a linear motor to control the distance between the drop centers $D$ through the speed of the substrate and the jetting frequency of the pipette. The time between the deposition was typically around 10~ms, which is much shorter than the relevant timescale over which the interaction is observed (from $t~=~0.28$~s onwards). The deposited drops were imaged from below the substrate using a camera (Ximea XiQ MQ013MG-ON) connected to a telecentric lens (Kowa LM50TC), and the experiment was illuminated from above (Schott Ace light source + diffuser plate). The spatial and temporal resolution were 3.5~$\mu$m/pixel and 20~ms, respectively. The images were processed to extract the time-dependent drop positions from which the separation distance $D$ and the interaction type (i.e. attraction or repulsion) were determined. 

%%%%% FIG 3
%%%%%%

%%%%%% FIG 4
\begin{figure}
\includegraphics{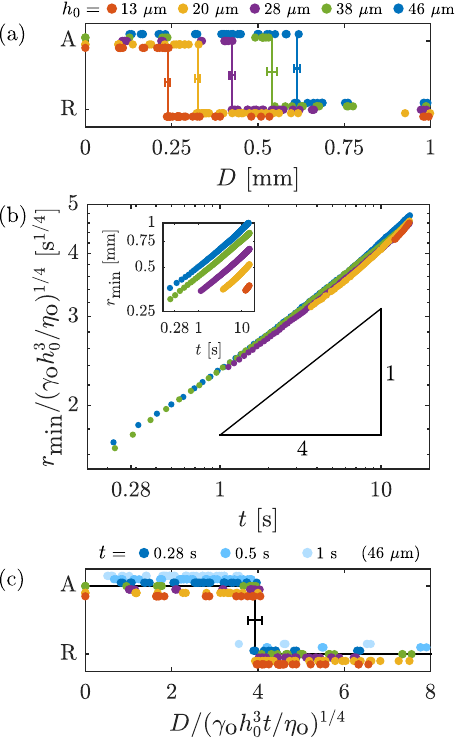}
\caption{\label{fig:fig4}(Color online) (a) Interaction type of two drops separated by a distance $D$ at $t = 0.28$~s, for films of varying thickness $h_0$. The solid lines indicate $D^\textrm{*}$, with the horizontal bars indicating the error in $D^\textrm{*}$. (b) The position of the dimple $r_\textrm{min}$ follows the scaling predicted by Eq.~(\ref{eq:Dscalinglaw}). Inset: unscaled data. (c) Same data as in (a) and Fig.~\ref{fig:fig3}a, but $D$ scaled according to Eq.~(\ref{eq:Dscalinglaw}). The solid black line shows the normalized value of $D^\textrm{*}$, which takes on a single value, the horizontal bar indicates the standard deviation.}
\end{figure}
%%%%%%

From the experiments we determine the type of interaction between two drops. In Fig.~\ref{fig:fig3} we show a typical series of experiments with  $h_0 = 46$~$\mu$m, varying the initial drop-drop separation distance $D$. In Fig.~\ref{fig:fig3}a we report whether the interaction is attractive (denoted A in the figure) or repulsive (denoted R) for varying distance $D$ and at various times $t$ after deposition. Here, each data point corresponds to one experiment with two drops. We observe a  sharp transition between attractive interactions (small $D$) and repulsive interactions (large $D$), and we denote the range of attractive interaction by the critical separation $D^\textrm{*}$.\footnote{We calculate $D^\textrm{*}$ by taking the mean of all $D$ values in the small region where attraction and repulsion overlap. The error in $D^\textrm{*}$ comprises of the standard deviation of these data points and the pixel error in $D$.} The experiments show that $D^\textrm{*}$ is not a universal length, as it observed to increase over time. Deformation of the drops, as observed from the top-view images, is small and occurs only when $D\ll D^\textrm{*}$ (i.e. in the case of attracting drops that are in very close proximity of each other), and thus does not affect the value of $D^\textrm{*}$. Note that the drops in Fig.~\ref{fig:fig1} remain circular in shape, except when in close proximity. Here we remark that $D^\textrm{*}$ is much larger than both $h_0$ and $R$; for example we measure $D^\textrm{*}~=~0.61$~mm at $t=0.28$~s. This observation is consistent with the behavior observed for the printed drop rows in Fig.~\ref{fig:fig1}, where the drop spacing is also the key factor that determines the interaction type. 

In a separate experiment, DHM was used to measure the drop-induced surface deformation of the oil film under the same conditions as for Fig.~\ref{fig:fig3}a. Fig.~\ref{fig:fig3}b shows the surface profile of the oil film from the center of the drop outward along a radial line with coordinate $r$, where $r=0$ corresponds to the drop center, at various times $t$. DHM is unable to measure the film profile close to the drop because the slope of the surface is too steep in this region (indicated by the black circle in Fig.~\ref{fig:fig2}). Comparison of Fig.~\ref{fig:fig3}a and Fig.~\ref{fig:fig3}b shows that $D^\textrm{*}$ is indeed directly comparable to the extent of the deformation of the film (approximately 0.1-1~mm), demonstrating that the interaction is indeed governed by this deformation. Since the deformation of the film evolves over time, the nature of the drop-drop interaction is time-dependent as well. 

Since the broadening of the surface deformations are expected to change with film thickness, we next study the dependence of $D^\textrm{*}$ on the film thickness $h_0$. The experiments from Fig.~\ref{fig:fig3}a were repeated with oil films of varying $h_0$, and the results are shown in Fig.~\ref{fig:fig4}a. Clearly, the value of $D^\textrm{*}$ strongly depends on $h_0$. Indeed, this can be correlated to the dynamics of the deformed surface, which exhibits a similar dependence on $h_0$. To demonstrate this, we characterize the film deformation for various film thicknesses using DHM. The time-evolution of the position of the dimple $r_\textrm{min}$, i.e. the first minimum in the profile as defined in Fig. \ref{fig:fig3}b, is plotted in the inset of Fig.~\ref{fig:fig4}b for several film thicknesses on a log-log scale. Clearly, the dynamics are strongly affected by the thickness of the film. This can be understood from the thin-film equation for the film profile $h(x,y)$, which reads:
\begin{equation}
\label{eq:thinfilmequation}
\frac{\partial h}{\partial t} = - \frac{\gamma_\textrm{o}}{3 \eta_\textrm{o}} \nabla \cdot \left( h^3 \nabla \nabla^2 h \right).
\end{equation}
The typical length-scale for the film thickness is $h_0$, while the gradient $\nabla$ acts along the lateral direction and is set by the radial distance to the drop $r$. With this, the terms in Eq.~(\ref{eq:thinfilmequation}) are expected to scale as:
\begin{equation}
\label{eq:Dscalinglaw}
\frac{h_0}{t} \propto \frac{\gamma_\textrm{o}}{\eta_\textrm{o}} \left(\frac{h_0^4}{r^4}\right) \quad \Rightarrow \quad r \propto \left( \frac{\gamma_\textrm{o}}{\eta_\textrm{o}}h_0^3 t \right)^{1/4}.
\end{equation}
Similar scaling laws have been observed for the flattening time of step-shaped thin polymer films\cite{Salez_PF_2012, McGraw_PRL_2012}, and for the wavelike deformation of a liquid close to a solid wall.\cite{Jensen_JEM_2004} The main panel of Fig.~\ref{fig:fig4}b shows the same data as the inset, rescaled using the scaling from Eq.~(\ref{eq:Dscalinglaw}), i.e. $r_\textrm{min}/(\gamma_\textrm{o}h_0^3/\eta_\textrm{o})^{1/4}$. The data collapses onto a universal curve, in agreement with Eq.~(\ref{eq:Dscalinglaw}).

We now apply the same scaling law to quantify the range of interaction between two drops. Fig.~\ref{fig:fig4}c shows the drop interaction type as a function of the drop spacing $D$ normalized using Eq.~(\ref{eq:Dscalinglaw}) for a range of film thicknesses. All data collapses on a single curve with the transition from attraction to repulsion at $D^\textrm{*}/(\gamma_\textrm{o} h_0^3 t/\eta_\textrm{o})^{1/4}~\approx~4$. Thus, the drop-induced deformation of the surface of the thin liquid film is indeed at the origin of the interactions. 

\begin{figure}
\includegraphics{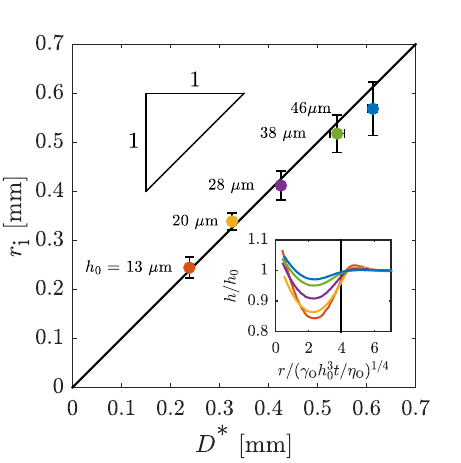}
\caption{\label{fig:fig5}(Color online) The distance at which the interaction transitions from attraction to repulsion $D^\textrm{*}$ as a function of the location of the inflection point of the surface profile $r_\textrm{i}$ at $t~=~0.28$~s. The solid black line indicates $D^\textrm{*} = r_\textrm{i}$ as a guide for the eye. Inset: deformation close to the drop on various film thicknesses normalized by the scaling law in Eq.~(\ref{eq:Dscalinglaw}). The solid black line corresponds to the transition from Fig.~\ref{fig:fig4}c.}
\end{figure}

Finally, we wish to quantify what property of the deformation determines whether drops attract or repel. In the example shown in Fig.~\ref{fig:fig3}, the transition between attraction and repulsion $D^*$ coincides with the inflection point of the deformed surface $r_\textrm{i}$.\footnote{We calculate $r_\textrm{i}$ by finding the $r$-coordinate of the maximum of $\partial h/\partial r$, which corresponds to $\partial^2 h/\partial r^2 = 0$. The error is determined by finding the two $r$-coordinates where $\partial h/\partial r$ is equal to $\partial h/\partial r - 0.05 \cdot \partial h/\partial r|_{r=r_\textrm{i}}$, where the last term is typical of the noise in the surface profile. Since the surface profiles flatten with increasing thickness, whereas the noise remains unchanged with thickness, the error bars increase for increasing $r_\textrm{i}$ in Fig.~\ref{fig:fig5}.} This is a general result, as can be seen from the inset of Fig. 5, where profiles of films with varying thickness have been rescaled according to the lubrication prediction. Indeed, in all cases, the critical distance  $D^\textrm{*}/(\gamma_\textrm{o} h_0^3 t/\eta_\textrm{o})^{1/4} \approx 4$ corresponds to the inflection point. This is further quantified in the main panel of Fig.~\ref{fig:fig5}, showing the direct correspondence of $D^\textrm{*}$ and $r_\textrm{i}$.\footnote{Due to the previously mentioned limited ability of DHM to resolve areas of the surface with high slopes we are unable to measure the deformation of the film at $t~=~0.28$~s for the three thinnest films ($h_0 = 13$, 20, 28~$\mu$m). We therefore measured at a time $t_\textrm{measure}~>~0.28$~s, and extrapolated back to $t = 0.28$~s using the scaling law from Eq. (\ref{eq:Dscalinglaw}) (in the form $r(t)~=~(t/t_\textrm{measure})^{1/4}r(t_\textrm{measure})$) to obtain the value of $r_\textrm{i}$ at $t~=~0.28$~s. Measurements of the films with $h_0~=~38$ and 46~$\mu$m were performed at $t = 0.28$~s and required no extrapolation.} Thus, we conclude that the interaction is determined by curvature of the viscous film. Intriguingly, this result appears to be different from the interaction between drops as observed on an elastic medium.\cite{Karpitschka_PNAS_2016, Pandey_SM_2017} In that case, the transition from attraction to repulsion was found to depend on whether the separation distance $D$ was small or large compared to the size of the drop $R$. For the case considered here, for which $D~\gg~R$, the elastic interaction can be described by a potential~$\sim \nabla^2 h$,\cite{Pandey_EL_2018} and the change from attraction to repulsion occurs when the potential has a maximum -- yet, in our experiments we find $D^\textrm{*}$ to occur when $\nabla^2 h\approx 0$ (i.e. not at its maximum), for reasons that remain to be identified. We emphasize once more that for the elastic case the interaction law does not change over time, while, by contrast, the viscous film evolves dynamically. These dynamics bring along additional viscous forces that may be the cause for the unexpected role of the inflection point of the profile.

%%%%%% FIG 5
%%%%%%

To summarize, we have observed non-monotonic capillary interactions between liquid drops on thin liquid films, focusing on the case of immiscible liquids. These non-monotonic interactions are due to visco-capillary waves on the viscous films, induced by the drops on the film. The range of the interaction is increasing with time, due to the broadening of the waves, which makes this ``viscous Cheerios effect" very different from the interactions observed on deep pools or on elastic substrates. Additionally, we have shown that the transition from attraction to repulsion coincides with the inflection point of the deformed surface. These results will be of importance for inkjet printing whenever drops are deposited on primer layers: capillary waves are also observed when drops are miscible, though in that case other factors such as mixing and Marangoni flows are expected to play a role. More generally, drop interactions on thin films might be of use for applications such as anti-fouling and self-assembly. For example for fog harvesting, substrates could be fine-tuned such that the interactions between drops lead to faster condensation of water.\cite{Sokuler_L_2010}
\\ \\
\emph{Supplementary Material} -- See supplementary material for movies of drop interaction. 
\\ \\
\indent This work is part of an Industrial Partnership Programme of the Foundation for Fundamental Research on Matter (FOM), which is financially supported by the Netherlands Organisation for Scientific Research (NWO). This research program is co-financed by Oc\'e-Technologies B.V., University of Twente, and Eindhoven University of Technology. S.K. acknowledges financial support from the University of Twente - Max Planck Center ``Complex fluid dynamics - Fluid dynamics of Complexity''. We also acknowledge support from D. Lohse's ERC Advanced Grant \#740479 -- DDD -- ERC-2016-ADG/ERC-2016-ADG out of which the DHM facility was financed. 

\nocite{*}
\bibliography{apl_wetonwet}

\end{document}